\documentclass[aps,pre,twocolumn,showpacs]{revtex4}
\usepackage{graphicx}
\usepackage{amssymb,amsfonts,amsmath}

\begin{document}

\title{Family name distributions: Master equation approach}
\author{Seung Ki Baek}
\author{Hoang Anh Tuan Kiet}
\author{Beom Jun Kim}
\email{beomjun@skku.edu}
\affiliation{Department of Physics, BK21 Physics Research
Division, and Institute of Basic Science, Sungkyunkwan University, Suwon
440-746, Republic of Korea}
\date{\today}

\begin{abstract}
Although cumulative family name distributions in many countries exhibit 
power-law forms, there also exist  counterexamples.
The origin of different family name distributions across countries
is discussed analytically in the framework of a population dynamics model.
Combined with empirical observations made, it is suggested that 
those differences in distributions are closely related to the 
rate of appearance of new family names.
\end{abstract}



\maketitle

\section{Introduction}
Understanding the structure of a population and how it evolves in 
time has been a critical issue in modern societies. 
Having started as an economic problem, it soon
extended to an environmental one, and states take censuses
periodically in order to use the results to design demographic policies.
Malthus was the first who presented the mathematical claim, which has been
accepted as the fundamental principle of population dynamics, that
``population, when unchecked, increases in a geometrical 
ratio''~\cite{Malthus}. In modern terms, he meant the exponential growth
tendency of a population of size $N$ with a constant net growth rate $r$,
i.e., $\dot N/N \equiv r$ with the time derivative $\dot N$;
this is often called the Malthusian growth model.
Forty years after Malthus published his essay, Verhulst added the
idea of the maximal capacity allowed by the environment, $K$, to the growth
model, so that the growth rate $r$ can be negative when the population
size $N$ exceeds $K$;
this is referred to as the logistic model~\cite{mathbiol}.
In addition to the inherent variety of dynamics it may exhibit
\cite{Strogatz}, there are also other models reflecting the complexity in
population dynamics, one of which has been introduced and termed
the $\theta$-logistic model~\cite{Saether, Sibly, Reynolds}.

Information on the structure of human population is not only 
available in many countries, but also very reliable owing to modern
census techniques. Various classifications therein allow deeper
insight into how subpopulations develop and interact with each other.
In this work, we classify people according to their family names 
(or surnames), and study the family name distribution. These studies
can also be important from the viewpoint
of genetics in biology, since the inheritance of the family name
is often paternal, exactly like the inheritance pattern of
the $Y$ chromosome. Furthermore, if
one can identify quantitatively the origin of differences in 
family name distributions across countries, it can provide an
understanding of the social mechanism behind the naming behavior
in human societies.

\begin{table}
\caption{Summary of the empirical results for family name distributions. 
The family name distribution function is written as 
$P(k) \sim k^{-\gamma}$  with $k$ the size of the family, which is the
number of individuals who have the same family name.
As the sampled population size $N$ is increased, the number $N_f$ of 
observed family names increases either logarithmically
(China and Korea) or algebraically (other countries), giving
us two distinct groups.
}
\begin{tabular*}{\hsize}{@{\extracolsep{\fill}}ccc}\hline\hline
Region      &$\gamma$   &$N_f$      \cr
\hline
China~\cite{Yuan}         &            &$\ln N$ \cr
Korea~\cite{BJKim}         & 1.0       &$\ln N$ \cr\cr
Argentina~\cite{Dipierri}     &        &$N^{0.84}$ \cr
Austria~\cite{Barrai}         &        &$N^{0.83}$ \cr
Berlin~\cite{Zanette}         &2.0        &\cr
France~\cite{Barrai}       &        &$N^{0.90}$ \cr
Germany~\cite{Barrai}         &        &$N^{0.77}$ \cr
Isle of Man~\cite{Reed}    &1.5       \cr
Italy~\cite{Barrai}        &        &$N^{0.75}$ \cr
Japan~\cite{Miyazima}         &1.75    &$N^{0.65}$ \cr
Netherlands~\cite{Barrai}     &        &$N^{0.69}$ \cr
Norway~\cite{ssb}        &2.16    &        \cr
Sicily~\cite{Pavesi}       &0.46--1.83  &$N^{1.0}$ \cr
Spain~\cite{Barrai}        &        &$N^{0.81}$ \cr
Switzerland~\cite{Barrai}     &        &$N^{0.73}$ \cr
Taiwan~\cite{Reed}         &1.9     &  \cr
United States~\cite{Zanette,Newman} &1.94    &     \cr
Venezuela~\cite{Alvaro}    &        &$N^{0.69}$ \cr
Vietnam~\cite{ho}         &1.43    &$N^{0.27}$ \cr
\hline\hline
\end{tabular*}
\label{table:summary}
\end{table}

The pattern of family name distributions in many countries 
have already been investigated (see Table~I):
In Japan, the family name distribution $P(k)$ has been shown to have
a power-law dependency on the size $k$ of families, i.e., 
$P(k)  \sim k^{-\gamma}$ with the 
exponent $\gamma \approx 1.75$~\cite{Miyazima}. Later, families
in the United States and Berlin have also been reported to display power-law
behavior with the similar exponent $\gamma \approx 2.0$ \cite{Zanette}. 
The same power-law distributions with exponents $\gamma \approx 1.9$
and $\gamma \approx 1.5$ have also been measured for 
Taiwanese family names and for names in the Isle of Man, 
respectively~\cite{Reed}.
Extensive research in various countries ranging from Western Europe to
South America has again found exponents around $2$ (See
Refs.~\cite{Barrai,Alvaro,Dipierri} and references therein).
In sharp contrast, the Korean family name
distribution has been recently investigated, revealing the very interesting
behavior of  $\gamma \approx 1.0$~\cite{BJKim}. The Korean distribution
is very different since the cumulative distribution $P_c(k)$
(the number of family names with more than $k$ members, divided by the total
number of family names)
becomes logarithmic, which results in an exponential Zipf plot
(sizes versus ranks of families)~\cite{BJKim}. Throughout the present
paper, the rank of a family is defined according to its size in 
descending order, i.e., the biggest family is assigned
rank 1, and the second biggest family has rank 2, and so on.
More strikingly, the exponentially decaying Zipf plot suggests 
that the number of family names $N_f$ found in a population of
size $N$ increases logarithmically in Korea (we observe
the same behavior for the Chinese family names reported in Ref.~\cite{Yuan}),
in sharp contrast 
to the corresponding results for other countries, where $N_f$ grows
algebraically with  $N$. 

In this work, we  investigate the possible mechanism for the
differences of family name distributions across countries, 
by using a simple model of population dynamics. We suggest that
the difference originates from the rate of appearance of
new family names, which is checked by empirical observation
made for the history of Korean family names. In more detail,
if new names appear linearly in time irrespective of the total
population size,  $\gamma=1$ is obtained, whereas if the number of new names 
generated per unit of time is proportional to the population size, $\gamma
\approx  2$ is concluded.
We also investigate the family books for several family names
in Korea containing genealogical trees, and extract the family name
distribution to construct the Zipf plot, revealing that
the exponential Zipf plot in Korea has been prevalent for at least
500 years. Family names in other countries such as 
China, Vietnam, and Norway are newly investigated, and comparisons
with existing studies lead  us to the conclusion that there are 
indeed two distinct groups of different family name distributions.

The present paper is organized as follows: We present our master equation
formulation, and obtain the formal solution for the distribution
function in Sec.~\ref{sec:formulation}. A detailed analysis is then
made in Sec.~\ref{sec:constant} for the case of constant
name generation rate, and historical observations are also discussed.
Section~\ref{sec:branching} is devoted to the other case when branching 
out from old to new names is allowed, which is followed by
a summary in Sec.~\ref{sec:summary}.

\section{Population dynamics: Formulation}
\label{sec:formulation}
We first introduce the master equation in a general
form to describe the time evolution of the family size, and then
present the formal solution obtained by using the generating function
technique. 

Let us define the probability $P_{j,k}(s,t)$ for a class (family)
to have number $n(t)=k$
at time $t$ given that it started with $n(s)=j$ at time $s$:
\begin{equation}
\label{eq:Pjkst}
P_{j,k}(s,t) = P[n(t)=k|n(s)=j],
\end{equation}
which is required to satisfy the initial condition
$P_{j,k}(s,s) = \delta_{jk}$ with the Kronecker delta $\delta_{jk} \equiv 1
(0)$ if $j = k$ ($j \neq k$).
The time evolution of $P_{j,k}(s,t)$ is governed by the following
master equation:
\begin{eqnarray}
\frac{dP_{j,k}(s,t)}{dt} &=& \lambda_{k-1}(t) P_{j,k-1}(s,t)\nonumber\\
& & + [\mu_{k+1}(t) + \beta_{k+1}(t)] P_{j,k+1}(s,t) \nonumber \\
& & - [\lambda_k(t) + \mu_k(t) + \beta_k(t)] P_{j,k}(s,t),
\label{eq:master}
\end{eqnarray}
where we have made the continuous-time approximation that
$P_{j,k}(s,t+1) - P_{j,k}(s,t) \approx dP_{j,k}(s,t)/dt$.
For convenience, we take one year as the time unit, and
thus the rate variables $\lambda$, $\mu$, and $\beta$ are
defined in terms of the annual change of population.
The first term in the right-hand side of Eq.~(\ref{eq:master}) describes
the process in which the class with $k-1$ members increases its members
by 1, which occurs at the birth rate $\lambda_{k-1}(t)$. 
The second term is for the opposite process that $k+1$ members
is decreased to $k$  members, which occurs when one member
either dies at the death rate $\mu_{k+1}(t)$, or invents a
new family name at the branching rate $\beta_{k+1}(t)$.  
We consider only the branching process in which a person invents
a new name; changing a name from one to an existing one
is not allowed in our model.
The last term is for the change from $k$ either  to $k+1$ or to $k-1$, which
occurs when a person is born, dies, and changes name,
at rates $\lambda_k(t)$, $\mu_k(t)$ and $\beta_k(t)$, respectively.
In this work, we  
allow birth and death rates to depend on time, and
write them as 
\begin{equation}
\label{eq:lambda}
\lambda_k(t) = k \lambda \phi(t) , \;
\mu_k(t) = k \mu \phi(t),
\beta_k(t) = k \beta \phi(t).
\end{equation}
The prefactor $k$ is easily understood since the family
with $k$ members has a chance proportional to $k$
to be picked up. We henceforth also assume that $\lambda > \mu + \beta$
to describe a population growing in time.

The solution of the master equation~[\ref{eq:master}] is easily found by using 
the generating function written as (see, e.g.,  Ref.~\cite{Parzen})
\begin{equation}
\label{eq:Psi}
\Psi_{j,s}(z,t) \equiv \sum_{k=0}^{\infty} z^k P_{j,k}(s,t), 
\end{equation}
with the initial condition $\Psi_{j,s}(z,s) =  z^j$ 
[see Eq.~(\ref{eq:Pjkst})].
It is straightforward to get the following partial differential equation
for $\Psi$ by combining Eqs.~(\ref{eq:master}) and (\ref{eq:Psi}):
\begin{equation*}
\frac{1}{\lambda \phi} \frac{\partial \Psi}{\partial t}
= (z-1) (z - \bar\mu) \frac{\partial \Psi}{\partial z} ,
\end{equation*}
with $\bar\mu \equiv (\mu+\beta)/\lambda < 1$, which is written in the simpler
form
\begin{equation*}
\frac{\partial \Psi}{\partial \tau} = \frac{\partial \Psi}{\partial x} ,
\end{equation*}
by introducing new variables $x$ and $\tau$ as
\begin{eqnarray*}
d \tau &\equiv& \lambda \phi ~dt,  \\
d x & \equiv & dz/(z-1) (z - \bar\mu).
\end{eqnarray*}
The solution should be written as $\Psi(x,\tau) = g(x+\tau)$, and the
functional form of $g(x)$ is determined from the initial condition
$\Psi_{j,s}(z,s) = z^j$ (we henceforth set $j=1$, i.e., the class started from 
only one member): 
\begin{eqnarray*}
\Psi = 1 + \left[ \left( \frac{1}{1-\bar\mu} + \frac{1}{z-1} \right)
e^{-(1-\bar\mu) (\tau - \sigma)} -\frac{1}{1-\bar\mu} \right]^{-1},
\end{eqnarray*}
where $\tau - \sigma = \int_s^t \lambda \phi(t') dt'$.
By expanding the generating function, we reach the desired probability
distribution
\begin{eqnarray*}
P_{1,k}(s,t) = \left\{
\begin{array}{lrl}
1-(1-\bar\mu)(1 - \bar\mu \eta)^{-1} & \mbox{for $k=0$,} \\
\eta^{k-1}(1-\eta)\left(1-\bar\mu\eta \right) & \mbox{for $k>0$,} 
\end{array}
\right.
\end{eqnarray*}
where
\begin{equation}
\label{eq:eta}
\eta  = \frac{1-e^{-(1-\bar\mu)(\tau - \sigma)}}{1-\bar\mu
	e^{-(1-\bar\mu)(\tau - \sigma)}}
= \frac{1-R}{1-\bar\mu R},
\end{equation}
with $R \equiv e^{-(1-\bar\mu)(\tau - \sigma)}$.

So far we have focused on the size of a class first introduced at a certain
time $s (<t)$. To derive the overall population distribution observed at time
$t$, one needs to know when each class was introduced. Let $\Pi(s)$
represent the rate at which a class is introduced. 
If the history begins at $t=0$, the resulting population distribution at
time $t$ is given by
\begin{equation}
P(k,t) = \frac{\int_{0}^{t} P_{1,k}(s,t) \Pi(s) ds}{\int_0^t \Pi(s)ds},
\label{eq:Pkt}
\end{equation}
where $N_f(t) \equiv\int_0^t \Pi(s) ds$ is the total number of family names 
at time $t$.

Although one can think of a further generalization using different 
time-dependent functions $\phi_\lambda(t)$, $\phi_\mu(t)$,  $\phi_\beta(t)$,  
for the corresponding  rate variables in Eq.~(\ref{eq:lambda}), for
simplicity we restrict 
ourselves only to the identical form $\phi(t)$. 
Within this limitation, 
it is noteworthy that our expression for the family name distribution
in Eq.~(\ref{eq:Pkt}) applies for a variety of different situations
for arbitrary $\Pi(s)$ and $\phi(t)$. For example, the widely used
Simon model~\cite{Zanette,Simon,Amaral,Newman,Yamasaki} corresponds to the
situation that $\Pi(s) = $ const and $\phi(t) \propto 1/N(t)$
with the total population $N(t)$. 
It is to be noted that the use
of $\phi(t) \propto 1/N(t)$ introduces an effective competition among
individuals: In one unit of time, only some fixed number of individuals are
allowed to
be born or die, which yields a linear increase of population in time,
different from what really happened in human history. 
Accordingly, we focus below on the case $\phi(t) = 1$ to have
exponential growth of the population; however, we consider different
choices for $\Pi(s)$.

The assumption of time-independent rates with $\phi(t)=1$ 
[see Eq.~(\ref{eq:lambda})] results in 
$(1-\bar\mu)(\tau - \sigma) = (\lambda -\mu -\beta) (t - s)$.
Without knowing the details of the generation mechanism of new family names,
it is  plausible to assume that new family names are introduced into
the population at the rate
\begin{equation*}
\Pi(s) = \alpha + \beta N(s), 
\end{equation*}
which contains both the population-independent part ($\alpha$) 
and the population-proportional part ($\beta N$).
The second term $\beta N(s)$ can be easily motivated if we assume that
each individual invents a new family name at a given probability $\beta$.
The population-independent part of the
name generation rate should also be included to describe, e.g., 
immigration from abroad.

Let us consider a family that started at time $s$. The expected family size
at time $t$ is computed to be [see Eq.~(\ref{eq:Psi})]
\begin{equation}
\label{eq:kst}
\bar k (s,t) = \sum_k k P_{1,k}(s,t) = 
\left.\frac{\partial \Psi}{\partial z}\right|_{z=1} = e^{(\lambda - \mu
-\beta)(t - s)}, 
\end{equation}
which yields the self-consistent integral equation
\begin{eqnarray*}
N(t) &=& \int_0^t \bar k(s,t) \Pi(s) ds, \\
	 &=& \int_0^t e^{(\lambda-\mu-\beta)(t-s)} [\alpha + \beta N(s)] ds,
\end{eqnarray*}
or, in the differential form,
\begin{equation*}
\frac{d N(t)}{dt} = \alpha + (\lambda-\mu) N(t),
\end{equation*}
with the solution 
\begin{equation}
\label{eq:Nt}
N(t) = N(0) e^{ (\lambda-\mu)t } + \frac{\alpha( e^{(\lambda-\mu)t} - 1)  }
         {\lambda-\mu}
     \propto e^{(\lambda-\mu)t}.
\end{equation}
As is expected, the population-proportional part $\beta N$ in $\Pi(s)$
due to the change of names (from the existing one to a new one) has nothing
to do with the increase or decrease of the total population, and thus only the 
population-independent part in $\Pi(s)$ enters $N(t)$.

\section{Constant name generation rate}
\label{sec:constant}
When family names appear uniformly in time and no branching process occurs,
i.e.,
\begin{equation*}
\Pi(s) = \alpha ,
\end{equation*}
we obtain, via change of the integration variable from $s$ to $\eta$
[see Eqs.~(\ref{eq:eta}) and (\ref{eq:Pkt})], 
\begin{equation*}
P(k,t) = \frac{1}{\lambda t} \int_0^1 \eta^{k-1} d\eta 
       = \frac{1}{\lambda t  k} \left( \frac{1 - e^{-(\lambda - \mu)t}}
           {1 - \bar\mu e^{-(\lambda - \mu)t}} \right)^k,
\end{equation*}
which yields
\begin{equation*}
P(k,t \rightarrow \infty) \sim k^{-\gamma}  \mbox{ ~with $\gamma=1$} .
\end{equation*}
It is also straightforward to get the number of family names 
\begin{equation}
\label{eq:Nf}
N_f(t) = \int_0^t \Pi(s) ds = \int_0^t \alpha~ ds \sim t,
\end{equation}
which, combined with the total population $N(t)  \sim e^{ (\lambda - \mu)t }$, 
yields the expression $N_f(t) \sim \ln N(t)$.

Interestingly, the above results are in  perfect agreement with what
has been found for Korean family name distribution~\cite{BJKim}. 
The cumulative family name distribution becomes logarithmic 
(i.e., $\gamma = 1$), which gives an exponentially decaying Zipf plot 
where the size is displayed as a function of the rank of the family.
Furthermore, one can show directly from $P(k) \sim k^{-1}$ that the 
number of family names $N_f$ found for the population size $N$
increases logarithmically, i.e., $N_f \sim \ln N$~\cite{BJKim}.

\begin{figure}
\includegraphics[width=3in]{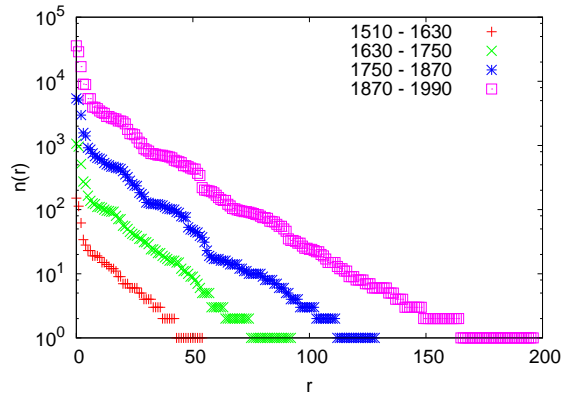}
\caption{(Color online) Korean family size $n(r)$ versus the 
rank $r$ of the family (Zipf plot) extracted from the family names of 
married women in family books. The exponential decay has been valid 
for at least 500 years in Korean history.}
\label{fig:kiet}
\end{figure}

The assumption of the constant rate $\Pi(s)$ of new name generation 
is  very plausible in Korean culture: More or less, it is 
considered as taboo to invent a new family name, and in Korean
history very few names have been introduced; there were only
288 family names in the year 2000~\cite{BJKim}. Furthermore, 
only 11 names newly appeared between 1985 and 2000, which seems to
imply that $\beta$ in Korea is extremely close to zero. (If we assume
that $\alpha = 0$, corresponding to the upper limit estimation 
of $\beta$, we get $\beta \approx 1.8\times 10^{-8}$ per year.)
Korea has preserved its family name system for more than two millennia, and
many Korean families still keep their own genealogical trees, 
from which their origins can be rather well dated. 
\begin{figure}
\includegraphics[width=3in]{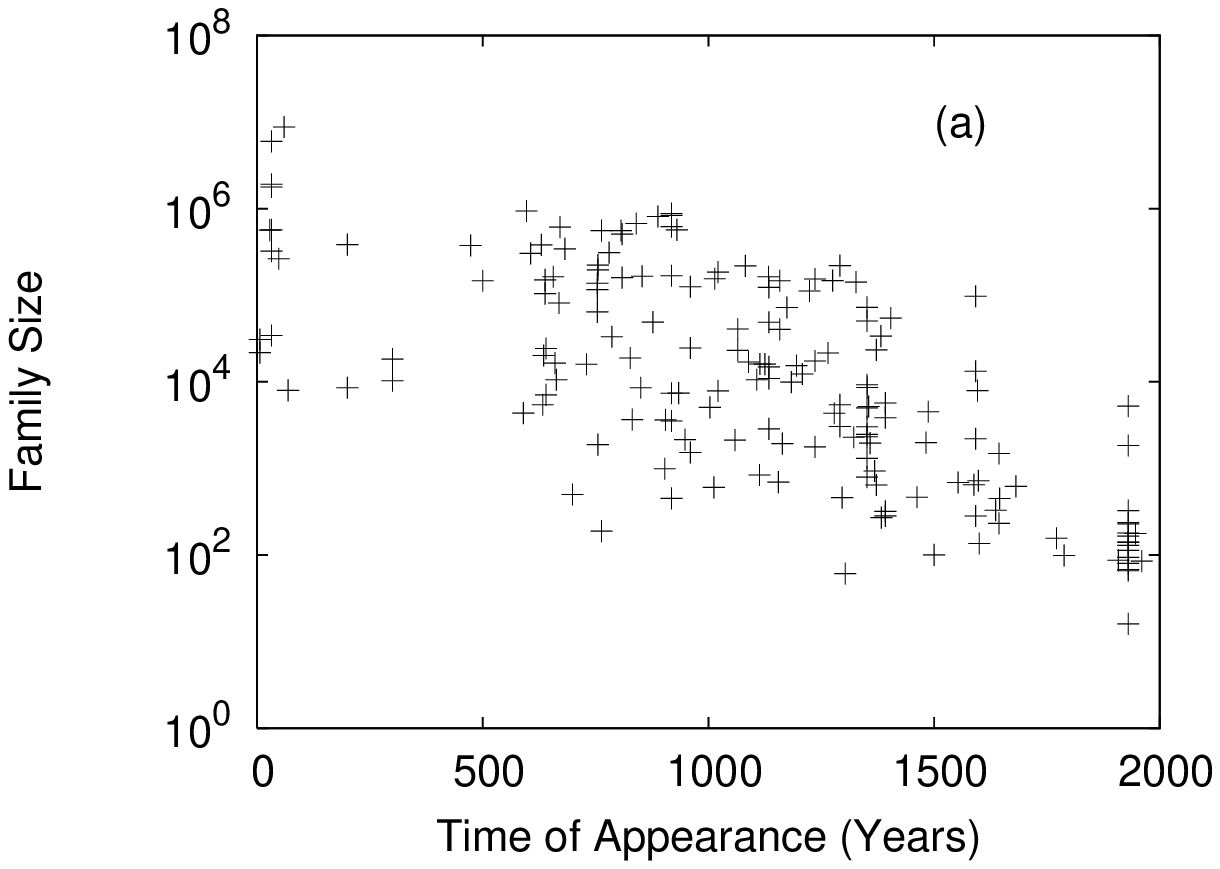}
\includegraphics[width=3in]{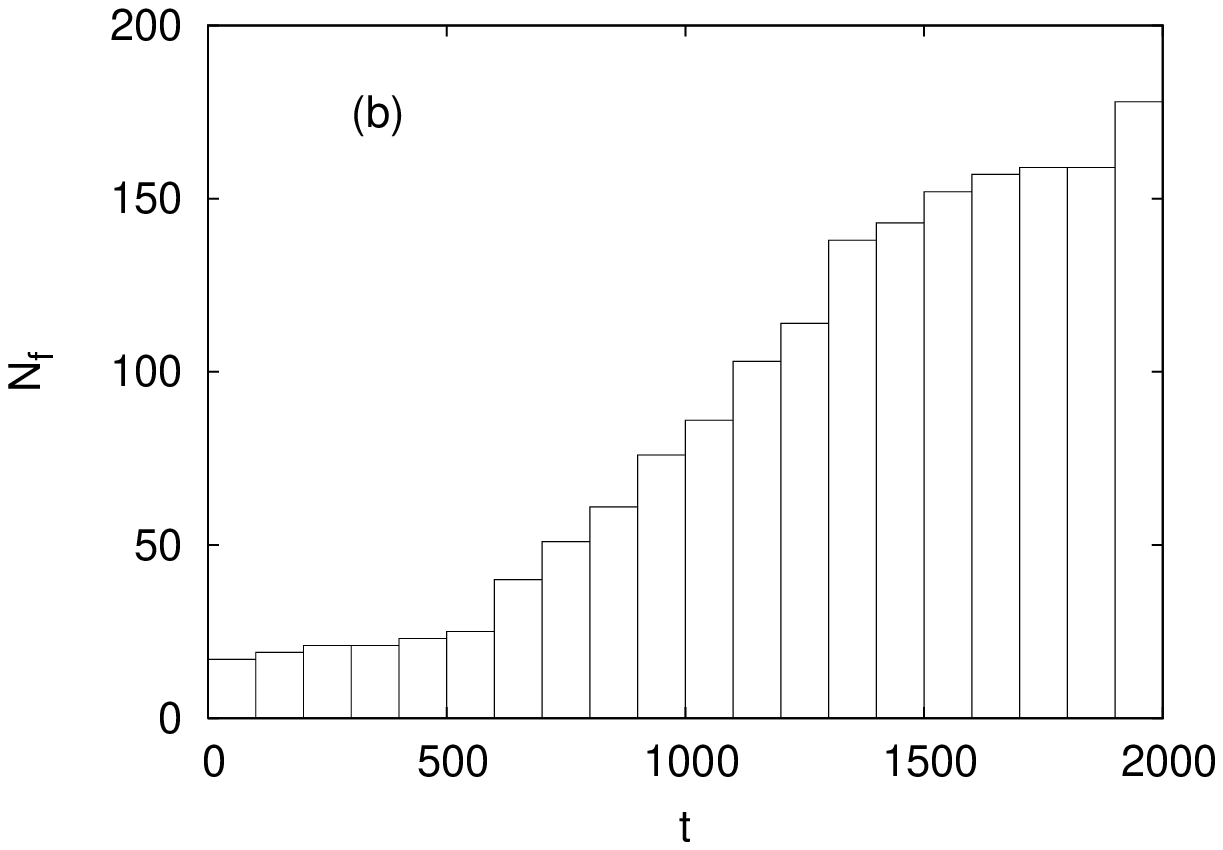}
\caption{(a) Each Korean family size versus its time of appearance.
The times are collected from the genealogical trees of 178 family
names, and the family sizes are from the governmental census data in 1985.
The family size grows exponentially since its first appearance as
time passes.
(b) Number of family names, $N_f$, as a function of time $t$ in units of
years. In Korea, $N_f$ grows approximately linearly in time while
the population growth is exponential.}
\label{fig:origin}
\end{figure}
In order to check the validity of the assumption of the
constant generation rate of names, we collect information about the
origins 
and sizes of family names from publicly accessible sources.
Collecting the sizes and times of appearance
for 178 family names, around 60\% of those existing, Fig.~\ref{fig:origin} is
obtained. In Fig.~\ref{fig:origin}(a), we show the present size of each family
versus the time when it first appeared and it seems  to be 
in accord with the exponential growth in Eq.~(\ref{eq:kst}). 
In Fig.~\ref{fig:origin}(b), we plot the number of family names
as a function of time. Although we have included only 178 names,
the plot is again in agreement with the linear increase of $N_f(t)$ in 
Eq.~(\ref{eq:Nf}) over a broad range of time. We emphasize here
that the number of Korean family names increases much more
slowly than the total population.
We also use several family books containing genealogical trees~\cite{kiet}.
Although these books contain only the paternal part of the trees,
the family names of women who were married to
the members of the family were recorded (in Korea, women do not
change their family names after their marriages). 
We use the information about family names
of women at various periods of time to plot Fig.~\ref{fig:kiet}. It is
clearly seen that the size of the family versus the
family rank decays exponentially for a broad range of periods, which
confirms that the exponential Zipf plot in Korea has been prevalent
for a long time and is not a recent trend.
We have shown above that a family name distribution of the form
$P(k) \sim k^{-1}$ is closely related to the constant generation rate
of new names, i.e., $\Pi(s) = \alpha$, which has also been validated 
from empirical historical observations.

\begin{figure}
\includegraphics[width=3in]{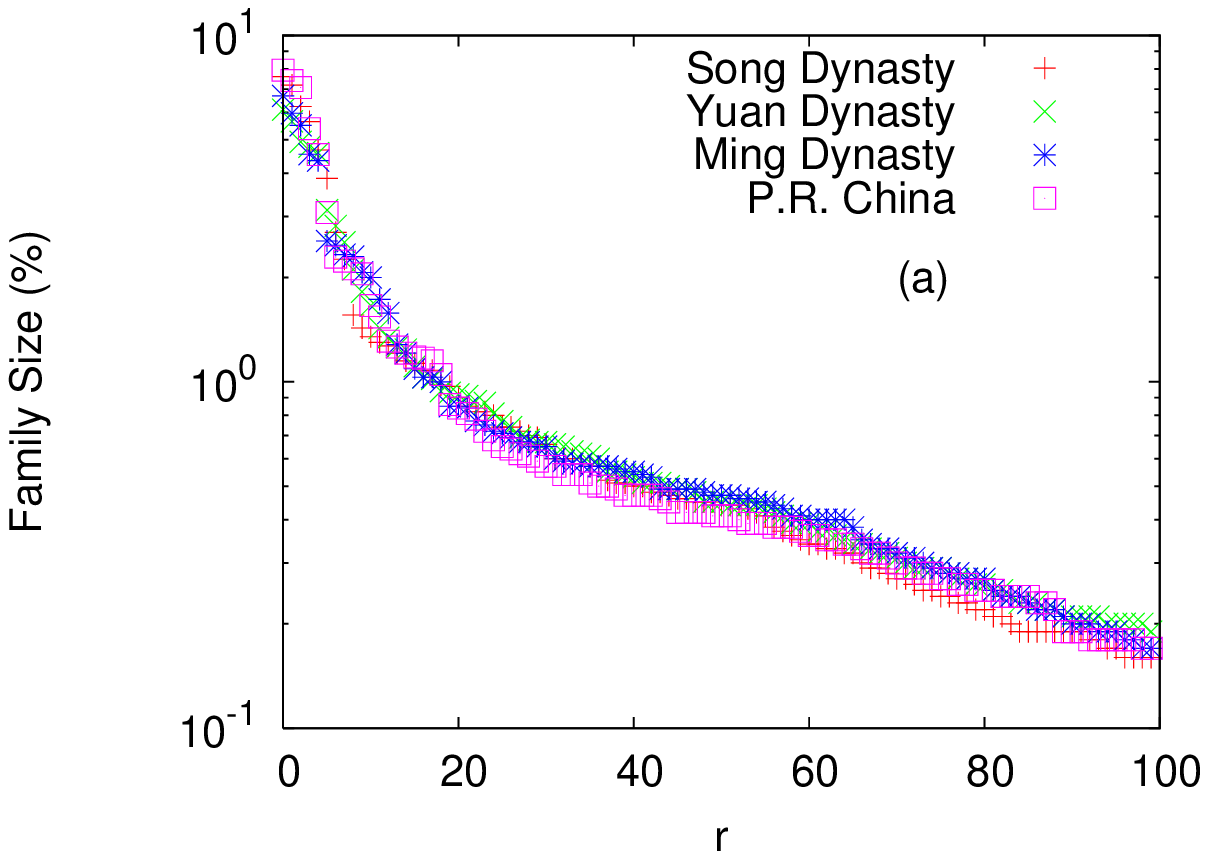}
\includegraphics[width=3in]{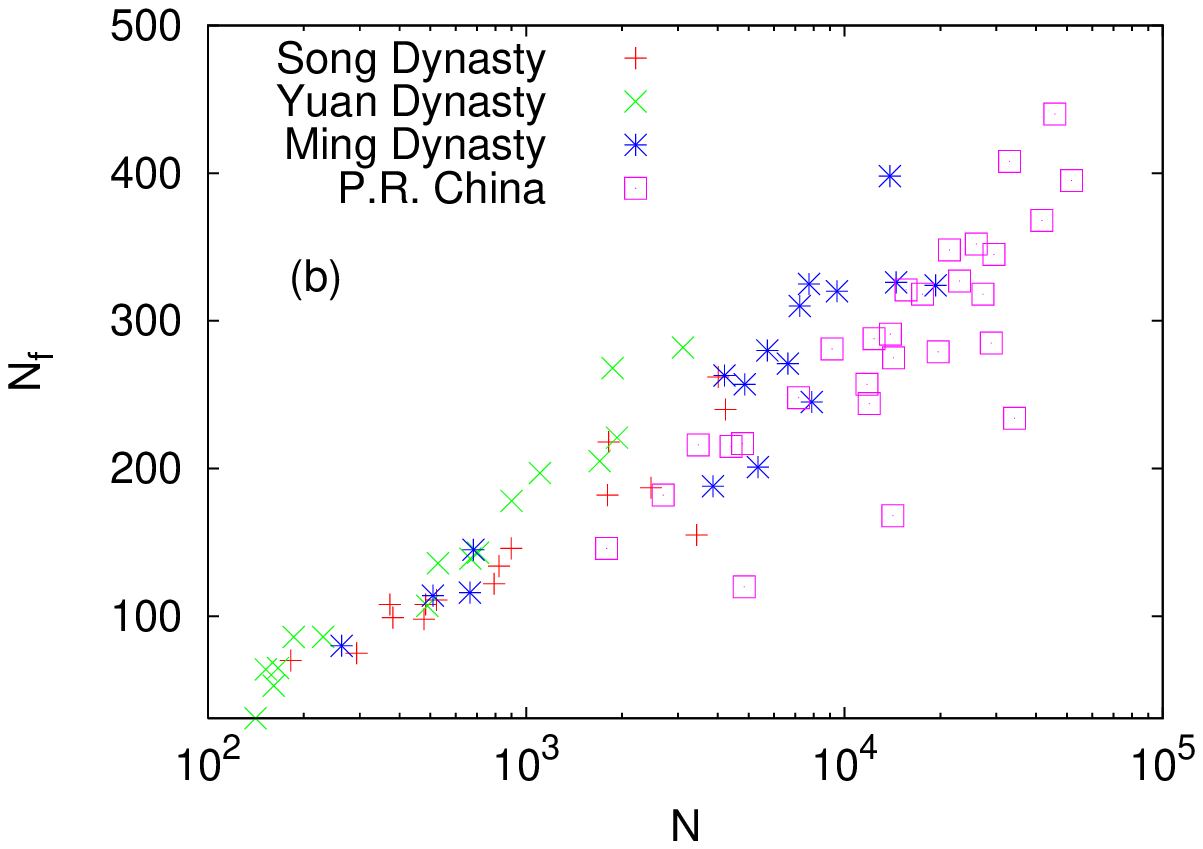}
\caption{(Color online) (a) Size distribution of Chinese family names,
arranged by their ranks $r$ (the Zipf plot).
The exponential shape has been maintained from the time of the Song Dynasty
(960--1279). (b) Number of people ($N$) versus the number of family names
found therein ($N_f$), collected in each province of China, showing clearly
$N_f \sim \ln N$.}
\label{fig:chinese}
\end{figure}

For another example, we present the result of our analysis
for Chinese family names in Ref.~\cite{Yuan}, where 
$542 262$ Chinese are sampled with $1042$ family names found. 
Although only the top $100$ Chinese family names are available in 
Ref.~\cite{Yuan}, the rank-size distribution (Zipf plot) appears to have
preserved 
an exponential tail for the almost a millennium, 
as shown in Fig.~\ref{fig:chinese}(a). Moreover, the number of
family names increases logarithmically with the number of people, as depicted
in Fig.~\ref{fig:chinese}(b), supporting our argument. 

We next pursue the answer to 
the question of how the distribution changes if new names are produced 
at a rate that is not fixed but grows with the population size.

\section{Family name distribution with branching process allowed}
\label{sec:branching}
If family names are allowed to branch out, the exponent $\gamma$
is altered. With $\beta$ being positive, $\Pi(s)$ is dominated by 
the exponential growth in the long run [see Eq.~(\ref{eq:Nt})]
\begin{equation*}
\Pi(s)	= \alpha + \beta N(s) \sim e^{(\lambda - \mu) s},
\end{equation*}
which we use to compute $P(k,t)$ in Eq.~(\ref{eq:Pkt}) as follows:
\begin{eqnarray*}
P(k,t \rightarrow \infty) &\propto& \int_{0}^{1} \eta^{k-1} e^{(\lambda-\mu)
	s} d\eta\\
	&\propto& k^{- \left\{ 1+(\lambda-\mu)/(\lambda-\mu-\beta)
		\right\}}.
\end{eqnarray*}
Consequently, the family name distribution in the case of $\beta > 0$ has 
\begin{equation*}
\gamma=2+\frac{\beta}{\lambda-\mu-\beta}  
\end{equation*}
in agreement with Ref.~\cite{Reed}. It is very reasonable to assume that
$\beta$ is much smaller than $\lambda - \mu$, and we expect
$\gamma \approx 2$ in most countries. Indeed, the United States
and Berlin have $\gamma \approx 2.0$~\cite{Zanette}, 
which, by using the relation
$N_f \sim N^{\gamma - 1}$ discussed in Ref.~\cite{BJKim}, leads to
the conclusion that $N_f$ and $N$ are proportional to each other.
Of course, one can confirm this linear relation from the direct 
calculation of the number of family names:
\begin{equation*}
N_f(t) = N_f(0) + \alpha \left( 1 - \frac{\beta}{\lambda - \mu} \right)
t + \frac{\beta}{\lambda-\mu} [N(t) - N(0)],
\end{equation*}
which confirms that $N_f/N \rightarrow \beta/(\lambda-\mu)$ 
as $t \rightarrow \infty$. 
The above result of $\gamma = 2 + \beta/(\lambda - \mu - \beta)$ should
be used carefully when $\beta \rightarrow 0$: If $\beta$ is strictly zero,
one cannot use the assumption
$\Pi(s) \sim e^{(\lambda - \mu)s}$, and we recover the result $\gamma = 1$
as previously shown.

From the publicly accessible population information, we estimate that the
Swedish population increased at the rate $\lambda - \mu \approx 0.456\%$ per
year during 2004--2006. In the same period of time, about 100 new
family names were introduced per month, which gives us a rough estimate $\beta
\approx 0.015 \%$. Accordingly, $\beta/(\lambda - \mu) \approx 0.03$,
which makes the assumption we made above very plausible. The number of family
names in Sweden is known to be somewhere between 140 000 and 400 000 depending
on how we count them. Together with the total population of about 9 millions,
we
confirm that $0.016 \lesssim (N_f/N) \lesssim 0.044$, which is in accord with
our expectation that $N_f/N = \beta/(\lambda - \mu) \approx 0.03$.

\begin{figure}
\includegraphics[width=3in]{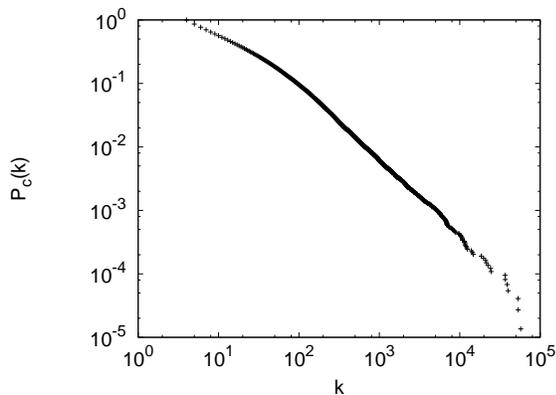}
\caption{Cumulative distribution $P_c(k)$ versus the family size $k$
of Norwegian family names, based on the survey in 2007~\cite{ssb}. 
The power-law behavior is clearly seen.}
\label{fig:norway}
\end{figure}

\begin{figure}
\includegraphics[width=3in]{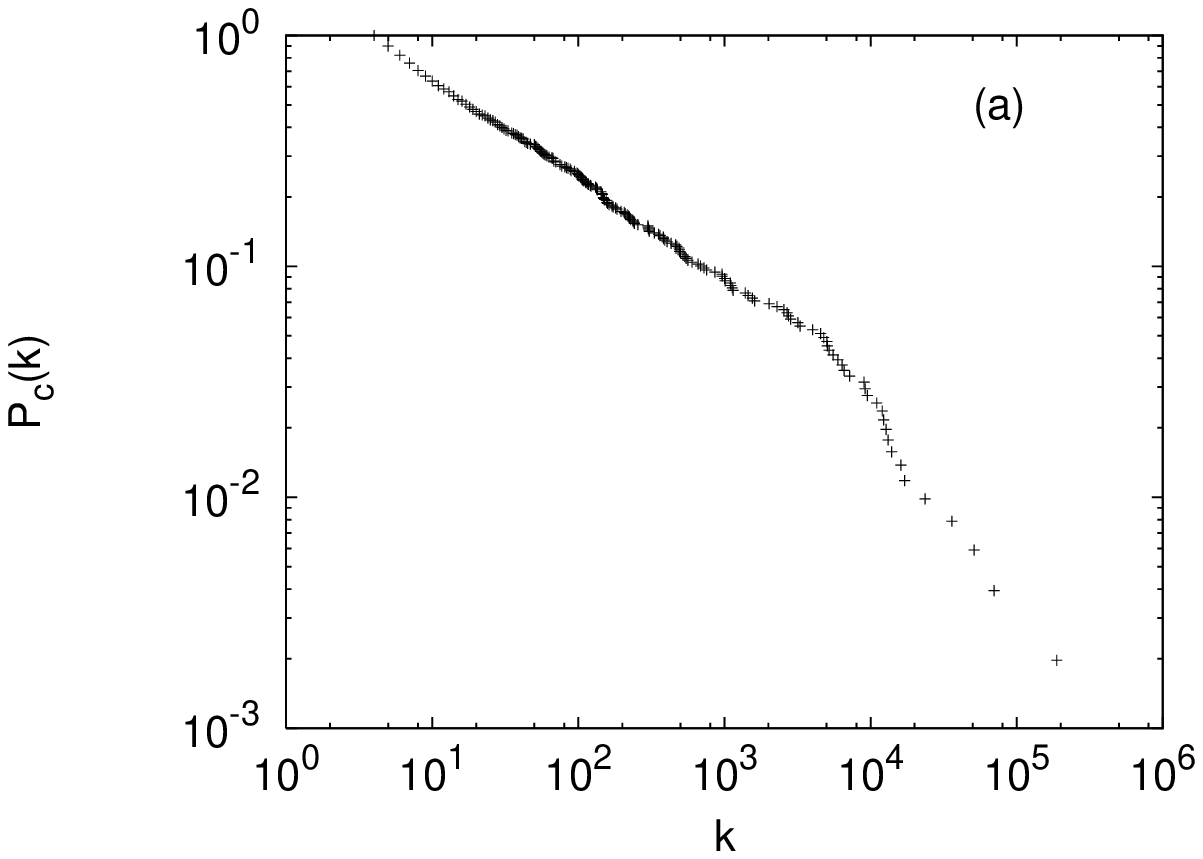}
\includegraphics[width=3in]{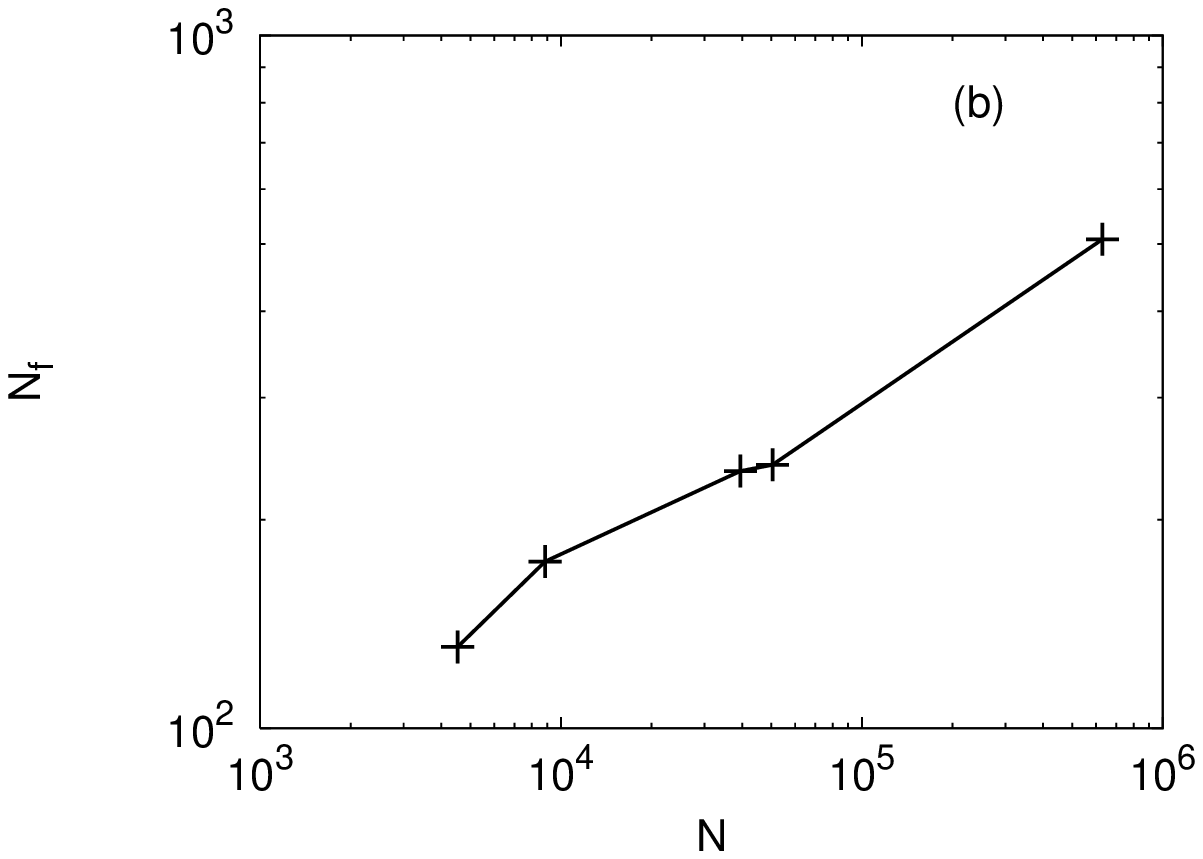}
\caption{(a) Vietnamese family name cumulative distribution, from the phone
book of Ho-Chi-Minh City, 2004. (b) The number of family names $N_f$ found in
certain numbers of people $N$. Both show power-law behaviors.
}
\label{fig:vietnam}
\end{figure}

The empirical findings we have referred to are listed in
Table~I, 
from which we suggest two main categories of family name systems: one 
with $\gamma \approx 1$ and a logarithmic increase of $N_f$ versus $N$
(Korea and China), and the other with $\gamma > 1$ and a power-law
increase of $N_f$ (other countries). The latter category has been 
prevalent in the literature, to which we also add Norway
with $\gamma \approx 2.16$ (Fig.~\ref{fig:norway}) and Vietnam
with $\gamma \approx 1.43$ (Fig.~\ref{fig:vietnam}).  

We suggest above that the existence of two groups of family
name distributions originates from the difference in new name
generation rates, which reflects the existence of a very different 
social dynamics  behind the naming behaviors across different cultures. 
We also point out that, due to the unavoidable simplifications made in our 
analytic model study, we are not able to clearly explain the 
spread of $\gamma$ in the second group of family names.
For example, in the history of Vietnam, when a dynasty was ruined
many Vietnamese belonging to the fallen dynasty changed their 
family names into existing ones, which our framework cannot take into
account.
Another interesting case is the Japanese system. Again, a Japanese family
name rarely undergoes the branching process these days, but
one finds the algebraic dependency of $N_f \sim
N^{0.65}$~\cite{Miyazima}, which indicates the fact that many Japanese
people had to adopt their family names by governmental policy about a
century ago. The diversity ensured at the creation appears to be
maintained up to now characterizing the Japanese family name system. 
Consequently, the Japanese name distribution cannot be successfully 
explained by our model in which the limit $t \rightarrow \infty$ 
is taken.
Another peculiar observation has been found in Sicily: The surname
distribution from one of its communes shows $\gamma \approx 0.46$,
possibly originating from the effects of isolation~\cite{Pavesi};
mathematical treatment of this population has not been carried out.
Within these limitations of our model study in which various simplifications
are made implicitly and explicitly, we strongly believe that
such an idealization in general helps one to sensitively check the reality 
and identify the most important issue from all the ingredients,
providing a deeper understanding and insight. 

\section{Summary}
\label{sec:summary}
In summary, we analytically investigated the generating
mechanism of observed family name distributions.
Whereas the traditional approaches from the Simon model are based on implicit
assumptions about competition within the population, we instead started
from the first principle of population dynamics, the Malthusian growth
model.
The existence of branching processes in generating new family names was
pointed out as the crucial factor in determining the power-law exponent
$\gamma$: With and without the branching process, $\gamma \approx 2$
and $\gamma  = 1$, respectively, were obtained.
Genealogical trees collected for Korean family names were analyzed
to confirm that the total number of names increased linearly in time,
justifying the assumption made in the analytic study.
We additionally reported Chinese, Vietnamese, and Norwegian
data sets to examine our argument, which, combined with existing
studies, lead us to the conclusion that there are two groups
of family name distributions on the globe and that these differences
can be successfully explained in terms of the differences in new name
generation rates.

\begin{acknowledgments}
We thank H. Jeong for providing us data from Korean family books, and
P. Holme for useful information on Swedish family names.
We also thank Statistics Norway for the Norwegian data.
This work was supported by the Korea Research Foundation Grant funded by the
Korean Government (MOEHRD), Grants No. KRF-2005-005-J11903 (S.K.B.),
No. KRF-2006-211-C00010 (H.A.T.K.), 
and No. KRF-2006-312-C00548 (B.J.K.).
\end{acknowledgments}


\begin{thebibliography}{20}
\expandafter\ifx\csname natexlab\endcsname\relax\def\natexlab#1{#1}\fi
\expandafter\ifx\csname bibnamefont\endcsname\relax
  \def\bibnamefont#1{#1}\fi
\expandafter\ifx\csname bibfnamefont\endcsname\relax
  \def\bibfnamefont#1{#1}\fi
\expandafter\ifx\csname citenamefont\endcsname\relax
  \def\citenamefont#1{#1}\fi
\expandafter\ifx\csname url\endcsname\relax
  \def\url#1{\texttt{#1}}\fi
\expandafter\ifx\csname urlprefix\endcsname\relax\def\urlprefix{URL }\fi
\providecommand{\bibinfo}[2]{#2}
\providecommand{\eprint}[2][]{\url{#2}}

\bibitem[{\citenamefont{Malthus}(1798)}]{Malthus}
\bibinfo{author}{\bibfnamefont{T.~R.} \bibnamefont{Malthus}},
  \emph{\bibinfo{title}{First Essay on Population}}
  (\bibinfo{publisher}{Macmillan \& Co.}, \bibinfo{address}{London},
  \bibinfo{year}{1798}).

\bibitem[{\citenamefont{Murray}(2003)}]{mathbiol}
\bibinfo{author}{\bibfnamefont{J.~D.} \bibnamefont{Murray}},
  \emph{\bibinfo{title}{Mathematical Biology}} (\bibinfo{publisher}{Springer},
  \bibinfo{address}{New York}, \bibinfo{year}{2003}), \bibinfo{edition}{3rd}
  ed.

\bibitem[{\citenamefont{Strogatz}(1994)}]{Strogatz}
\bibinfo{author}{\bibfnamefont{S.}~\bibnamefont{Strogatz}},
  \emph{\bibinfo{title}{Nonlinear Dynamics and Chaos}}
  (\bibinfo{publisher}{Addison-Wesley}, \bibinfo{address}{Reading,
  MA}, \bibinfo{year}{1994}).

\bibitem[{\citenamefont{S{\ae}ther et~al.}(2002)\citenamefont{S{\ae}ther,
  Engen, and Matthysen}}]{Saether}
\bibinfo{author}{\bibfnamefont{B.-E.} \bibnamefont{S{\ae}ther}},
  \bibinfo{author}{\bibfnamefont{S.}~\bibnamefont{Engen}}, \bibnamefont{and}
  \bibinfo{author}{\bibfnamefont{E.}~\bibnamefont{Matthysen}},
  \bibinfo{journal}{Science} \textbf{\bibinfo{volume}{295}},
  \bibinfo{pages}{2070} (\bibinfo{year}{2002}).

\bibitem[{\citenamefont{Sibly et~al.}(2005)\citenamefont{Sibly, Barker, Denham,
  Hone, and Pagel}}]{Sibly}
\bibinfo{author}{\bibfnamefont{R.~M.} \bibnamefont{Sibly}},
  \bibinfo{author}{\bibfnamefont{D.}~\bibnamefont{Barker}},
  \bibinfo{author}{\bibfnamefont{M.~C.} \bibnamefont{Denham}},
  \bibinfo{author}{\bibfnamefont{J.}~\bibnamefont{Hone}}, \bibnamefont{and}
  \bibinfo{author}{\bibfnamefont{M.}~\bibnamefont{Pagel}},
  \bibinfo{journal}{Science} \textbf{\bibinfo{volume}{309}},
  \bibinfo{pages}{607} (\bibinfo{year}{2005}).

\bibitem[{\citenamefont{Reynolds and Freckleton}(2005)}]{Reynolds}
\bibinfo{author}{\bibfnamefont{J.~D.} \bibnamefont{Reynolds}} \bibnamefont{and}
  \bibinfo{author}{\bibfnamefont{R.~P.} \bibnamefont{Freckleton}},
  \bibinfo{journal}{Science} \textbf{\bibinfo{volume}{309}},
  \bibinfo{pages}{567} (\bibinfo{year}{2005}).

\bibitem[{\citenamefont{Yuan}(2002)}]{Yuan}
\bibinfo{author}{\bibfnamefont{Y.}~\bibnamefont{Yuan}},
  \emph{\bibinfo{title}{Chinese Surnames}}
  (\bibinfo{publisher}{East China Normal University Press},
  \bibinfo{address}{Sanghai}, \bibinfo{year}{2002}) (in Chinese).

\bibitem[{\citenamefont{Kim and Park}(2005)}]{BJKim}
\bibinfo{author}{\bibfnamefont{B.~J.} \bibnamefont{Kim}} \bibnamefont{and}
  \bibinfo{author}{\bibfnamefont{S.~M.} \bibnamefont{Park}},
  \bibinfo{journal}{Physica A} \textbf{\bibinfo{volume}{347}},
  \bibinfo{pages}{683} (\bibinfo{year}{2005}).

\bibitem[{\citenamefont{Dipierri et~al.}(2005)\citenamefont{Dipierri, Alfaro,
  Scapoli, Mamolini, Rodriguez-Larralde, and Barrai}}]{Dipierri}
\bibinfo{author}{\bibfnamefont{J.~E.} \bibnamefont{Dipierri}},
  \bibinfo{author}{\bibfnamefont{E.~L.} \bibnamefont{Alfaro}},
  \bibinfo{author}{\bibfnamefont{C.}~\bibnamefont{Scapoli}},
  \bibinfo{author}{\bibfnamefont{E.}~\bibnamefont{Mamolini}},
  \bibinfo{author}{\bibfnamefont{A.}~\bibnamefont{Rodriguez-Larralde}},
  \bibnamefont{and} \bibinfo{author}{\bibfnamefont{I.}~\bibnamefont{Barrai}},
  \bibinfo{journal}{Am. J. Phys. Anthropol.} \textbf{\bibinfo{volume}{128}},
  \bibinfo{pages}{199} (\bibinfo{year}{2005}).

\bibitem[{\citenamefont{Scapoli et~al.}(2007)\citenamefont{Scapoli, Mamolini,
  Carrieri, Rodriguez-Larralde, and Barrai}}]{Barrai}
\bibinfo{author}{\bibfnamefont{C.}~\bibnamefont{Scapoli}},
  \bibinfo{author}{\bibfnamefont{E.}~\bibnamefont{Mamolini}},
  \bibinfo{author}{\bibfnamefont{A.}~\bibnamefont{Carrieri}},
  \bibinfo{author}{\bibfnamefont{A.}~\bibnamefont{Rodriguez-Larralde}},
  \bibnamefont{and} \bibinfo{author}{\bibfnamefont{I.}~\bibnamefont{Barrai}},
  \bibinfo{journal}{Theor. Popul. Biol.} \textbf{\bibinfo{volume}{71}},
  \bibinfo{pages}{37} (\bibinfo{year}{2007}).

\bibitem[{\citenamefont{Zanette and Manrubia}(2001)}]{Zanette}
\bibinfo{author}{\bibfnamefont{D.~H.} \bibnamefont{Zanette}} \bibnamefont{and}
  \bibinfo{author}{\bibfnamefont{S.~C.} \bibnamefont{Manrubia}},
  \bibinfo{journal}{Physica A} \textbf{\bibinfo{volume}{295}},
  \bibinfo{pages}{1} (\bibinfo{year}{2001}).

\bibitem[{\citenamefont{Reed and Hughes}(2003)}]{Reed}
\bibinfo{author}{\bibfnamefont{W.~J.} \bibnamefont{Reed}} \bibnamefont{and}
  \bibinfo{author}{\bibfnamefont{B.~D.} \bibnamefont{Hughes}},
  \bibinfo{journal}{Physica A} \textbf{\bibinfo{volume}{319}},
  \bibinfo{pages}{579} (\bibinfo{year}{2003}).

\bibitem[{\citenamefont{Miyazima et~al.}(2000)\citenamefont{Miyazima, Lee,
  Nagamine, and Miyajima}}]{Miyazima}
\bibinfo{author}{\bibfnamefont{S.}~\bibnamefont{Miyazima}},
  \bibinfo{author}{\bibfnamefont{Y.}~\bibnamefont{Lee}},
  \bibinfo{author}{\bibfnamefont{T.}~\bibnamefont{Nagamine}}, \bibnamefont{and}
  \bibinfo{author}{\bibfnamefont{H.}~\bibnamefont{Miyajima}},
  \bibinfo{journal}{Physica A} \textbf{\bibinfo{volume}{278}},
  \bibinfo{pages}{282} (\bibinfo{year}{2000}).

\bibitem[{\citenamefont{Pavesi et~al.}(2003)\citenamefont{Pavesi, Pizzetti,
  Siri, Lucchetti, and Conterio}}]{Pavesi}
\bibinfo{author}{\bibfnamefont{A.}~\bibnamefont{Pavesi}},
  \bibinfo{author}{\bibfnamefont{P.}~\bibnamefont{Pizzetti}},
  \bibinfo{author}{\bibfnamefont{E.}~\bibnamefont{Siri}},
  \bibinfo{author}{\bibfnamefont{E.}~\bibnamefont{Lucchetti}},
  \bibnamefont{and} \bibinfo{author}{\bibfnamefont{F.}~\bibnamefont{Conterio}},
  \bibinfo{journal}{Am. J. Phys. Anthropol.} \textbf{\bibinfo{volume}{120}},
  \bibinfo{pages}{195} (\bibinfo{year}{2003}).

\bibitem[{\citenamefont{Newman}(2005)}]{Newman}
\bibinfo{author}{\bibfnamefont{M.~E.~J.} \bibnamefont{Newman}},
  \bibinfo{journal}{Contemp. Phys.} \textbf{\bibinfo{volume}{46}},
  \bibinfo{pages}{323} (\bibinfo{year}{2005}).

\bibitem[{\citenamefont{Rodriguez-Larralde
  et~al.}(2000)\citenamefont{Rodriguez-Larralde, Morales, and Barrai}}]{Alvaro}
\bibinfo{author}{\bibfnamefont{A.}~\bibnamefont{Rodriguez-Larralde}},
  \bibinfo{author}{\bibfnamefont{J.}~\bibnamefont{Morales}}, \bibnamefont{and}
  \bibinfo{author}{\bibfnamefont{I.}~\bibnamefont{Barrai}},
  \bibinfo{journal}{Am. J. Hum. Biol.} \textbf{\bibinfo{volume}{12}},
  \bibinfo{pages}{352} (\bibinfo{year}{2000}).

\bibitem[{\citenamefont{Parzen}(1962)}]{Parzen}
\bibinfo{author}{\bibfnamefont{E.}~\bibnamefont{Parzen}},
  \emph{\bibinfo{title}{Stochastic Processes}}
  (\bibinfo{publisher}{Holden-Day}, \bibinfo{address}{San Francisco},
  \bibinfo{year}{1962}).

\bibitem[{\citenamefont{Simon}(1955)}]{Simon}
\bibinfo{author}{\bibfnamefont{H.~A.} \bibnamefont{Simon}},
  \bibinfo{journal}{Biometrika} \textbf{\bibinfo{volume}{42}},
  \bibinfo{pages}{425} (\bibinfo{year}{1955}).

\bibitem[{\citenamefont{Amaral et~al.}(2000)\citenamefont{Amaral, Scala,
  Barth\'el\'emy, and Stanley}}]{Amaral}
\bibinfo{author}{\bibfnamefont{L.~A.~N.} \bibnamefont{Amaral}},
  \bibinfo{author}{\bibfnamefont{A.}~\bibnamefont{Scala}},
  \bibinfo{author}{\bibfnamefont{M.}~\bibnamefont{Barth\'el\'emy}},
  \bibnamefont{and} \bibinfo{author}{\bibfnamefont{H.~E.}
  \bibnamefont{Stanley}}, \bibinfo{journal}{Proc. Nat. Acad. Sci. U.S.A.}
  \textbf{\bibinfo{volume}{97}}, \bibinfo{pages}{11149} (\bibinfo{year}{2000}).

\bibitem[{\citenamefont{Yamasaki et~al.}(2006)\citenamefont{Yamasaki, Matia,
  Buldyrev, Fu, Pammolli, Riccaboni, and Stanley}}]{Yamasaki}
\bibinfo{author}{\bibfnamefont{K.}~\bibnamefont{Yamasaki}},
  \bibinfo{author}{\bibfnamefont{K.}~\bibnamefont{Matia}},
  \bibinfo{author}{\bibfnamefont{S.~V.} \bibnamefont{Buldyrev}},
  \bibinfo{author}{\bibfnamefont{D.}~\bibnamefont{Fu}},
  \bibinfo{author}{\bibfnamefont{F.}~\bibnamefont{Pammolli}},
  \bibinfo{author}{\bibfnamefont{M.}~\bibnamefont{Riccaboni}},
  \bibnamefont{and} \bibinfo{author}{\bibfnamefont{H.~E.}
  \bibnamefont{Stanley}}, \bibinfo{journal}{Phys. Rev. E}
  \textbf{\bibinfo{volume}{74}}, \bibinfo{pages}{035103(R)}
  (\bibinfo{year}{2006}).

\bibitem{kiet} H.A.T. Kiet, S.K. Baek, and H. Jeong, and B.J. Kim, 
J. Korean Phys. Soc. \textbf{51}, 1812 (2007).

\bibitem{ssb} Statistics Norway (http://ssb.no).
\bibitem{ho} Telephone Directory 2004, Ho-Chi-Minh City, Vietnam.

\end{thebibliography}

\end{document}